\documentclass[aps,preprint,superscriptaddress,groupedaddress]{revtex4}  
\usepackage{graphicx}  
\usepackage{dcolumn}   
\usepackage{bm}        
\usepackage{amssymb}   
\usepackage{amsmath}   

\begin{document}


\hspace{5.2in} 

\title{Improved coarse-graining of Markov state models via explicit consideration of statistical uncertainty}
\author{Gregory R. Bowman}
\affiliation{Departments of Chemistry and Molecular \& Cell Biology, University of California, Berkeley, CA, 94720}
\email{gregoryrbowman@gmail.com}
\date{\today}

\begin{abstract}
Markov state models (MSMs)---or discrete-time master equation models---are a powerful way of modeling the structure and function of molecular systems like proteins.
Unfortunately, MSMs with sufficiently many states to make a quantitative connection with experiments (often tens of thousands of states even for small systems) are generally too complicated to understand. 
Here, I present a Bayesian agglomerative clustering engine (BACE) for coarse-graining such Markov models, thereby reducing their complexity and making them more comprehensible.
An important feature of this algorithm is its ability to explicitly account for statistical uncertainty in model parameters that arises from finite sampling.
This advance builds on a number of recent works highlighting the importance of accounting for uncertainty in the analysis of MSMs and provides significant advantages over existing methods for coarse-graining Markov state models.
The closed-form expression I derive here for determining which states to merge is equivalent to the generalized Jensen-Shannon divergence, an important measure from information theory that is related to the relative entropy.
Therefore, the method has an appealing information theoretic interpretation in terms of minimizing information loss.
The bottom-up nature of the algorithm likely makes it particularly well suited for constructing mesoscale models.
I also present an extremely efficient expression for Bayesian model comparison that can be used to identify the most meaningful levels of the hierarchy of models from BACE.
\end{abstract}

\maketitle


\section{Introduction}

Markov state models (MSMs) are a powerful means of understanding dynamic processes on the molecular scale, like protein folding and function \cite{BOWMAN:2010p1,prinz,pavel}.  
These discrete-time master equation models consist of a set of statesÑ--akin to local minima in the system's free energy landscapeÑ--and a matrix of transition probabilities between them.
Typically, the states are identified via a kinetic clustering of molecular dynamics simulations.

Unfortunately, building MSMs and extracting understanding from them is still a challenging task.
Ideally, MSMs would be constructed using a purely kinetic clustering of a simulation data set.
Calculating the transition rate between two conformations is an unsolved problem though, so, a number of alternative methods for building MSMs have been developed \cite{chodera, hierarchical, msmb, rains, keller, emma}.
Many of these approaches have converged on a two-stage process.
First, the conformations sampled are clustered into microstates based on geometric criteria such that the degree of geometric similarity between conformations in the same state implies a kinetic similarity.
Such models are excellent for making a quantitative connection with experiments because of their high temporal and spatial resolution.
However, it is difficult to examine such models to gain an intuition for a system because the rugged nature of most biomoleculeÕs free energy landscapes requires that the initial microstate model have tens of thousands of states.
Therefore, in a second stage, the initial state space is coarse-grained by lumping rapidly interconvertingÑ--or kinetically close--Ñmicrostates together into macrostates to obtain a more compact and comprehensible model.
Reasonable methods are now available for the first stage of this procedure \cite{chodera, hierarchical, msmb, rains, keller, emma}, but there is still a need for more efficient and accurate methods for coarse-graining MSMs.

\begin{figure}
\includegraphics[scale=0.8]{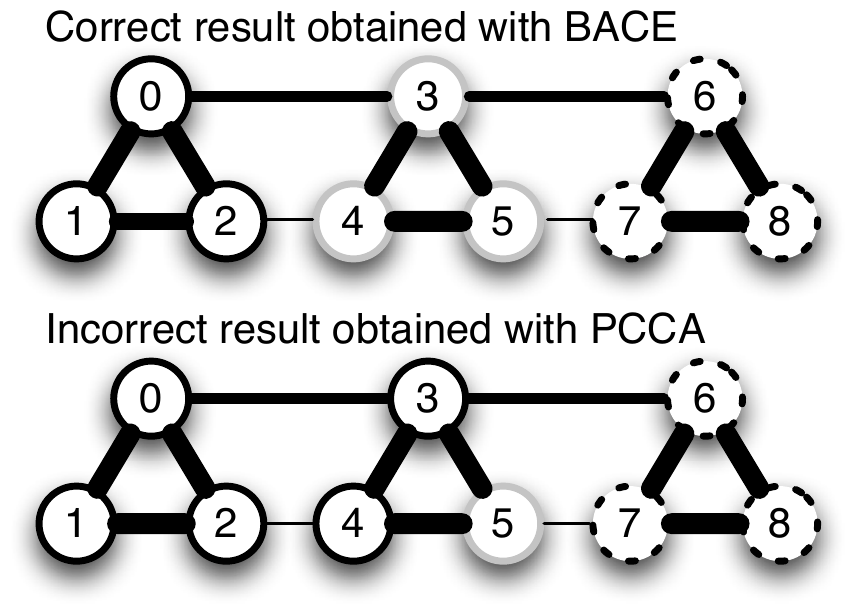}
\caption{\label{fig:simpleModel} A simple model demonstrating that BACE correctly deals with poorly sampled transitions, whereas eigenspectrum-based methods like PCCA are confounded by them.
This simple model has nine microstates (circles) whose borders are colored (solid black, gray, or dashed black) according to their assignment to three macrostates using either BACE or PCCA.
Each microstate has 1000 self-transitions, thick connections represent 100 transitions, medium lines represent 10 transitions, and thin lines represent 1 transition.
Therefore, the best coarse-graining into three states is to merge states 0-2, 3-5, and 6-8 because transitions within these groups are fast compared to transitions between the groups.
BACE correctly identifies this optimal coarse-graining into three macrostates.
However, the poorly sampled transitions between states 2-4 and 5-7 cause PCCA to mistakenly assign states 3-4 with stets 0-2 instead of with state 5.
}
\end{figure}

A major challenge in coarse-graining MSMs is dealing with uncertainty.
The most common methods for coarse-graining MSMs are Perron Cluster Cluster Analysis (PCCA) \cite{PCCA,PCCA2} and PCCA+ \cite{PCCAPlus}, though a number of new methods have been published recently \cite{rains, Huang:2010p22, Biswas, Vitalis, stock}.
Most all of these methods operate on the maximum-likelihood estimate of the transition probability matrix and do not account for statistical uncertainty in these parameters due to finite sampling.
For example, PCCA(+) uses the eigenspectrum of the transition matrix to find the partitioning that best captures the slowest transitions.
Such methods are well suited to data-rich situations but often fail when poorly sampled transitions are present \cite{Huang:2010p22}.
For example, Fig. \ref{fig:simpleModel} shows a case where PCCA fails due to a few poorly sampled transitions.
Eigenspectrum-based methods also have trouble creating mesoscale models---models with a large number of macrostates that are still quantitatively predictive yet are significantly more compact than the original microstate model---due to issues like propagating error.

Here, I present a Bayesian agglomerative clustering engine (BACE) for coarse-graining MSMs in a manner that accounts for model uncertainty and can easily create mesoscale models.
Bayesian methods have found wide applications in the physical sciences, and in MSMs in particular \cite{Hinrichs:2007p1865,NoeSampling,Sriraman:2005p3010,Bacallado:2009p1009}, for their ability to deal with uncertainty.
Inspired by the hierarchical nature of biomolecules' free energy landscapes, BACE performs an agglomerative clustering of microstates into macrostates by iteratively lumping together the most kinetically similar states---i.e. the most rapidly mixing states.
The key equation derived here is a closed-form expression for a Bayes factor that quantifies how likely two states are to be kinetically identical.
This expression is related to the relative entropy \cite{RelEnt}, an information theoretic measure that has found numerous applications in the physical sciences \cite{Shell:2008p3015,BOWMAN:2010p17,Crooks}.
Indeed, the expression is actually equivalent to the generalized Jensen-Shannon divergence \cite{Jensen}, which can be interpreted as the average amount of information a single measurement gives about which of two possible distributions is being sampled.
I also present an approximate expression for model comparison that allows one to identify the most informative levels of the hierarchy of models generated with BACE.
These methods could be applied directly to other Markov processes and could also be extended to other probabilistic models.

Code is available on the web (https://sites.google.com/site/gregoryrbowman/) and through the msmbuilder project (https://simtk.org/home/msmbuilder) \cite{msmb,Beauchamp:2012p2556}.

\section{BACE Algorithm}
The hierarchical structure of biomolecules' free energy landscapes naturally suggests a hierarchical approach to model construction.
The free energy landscapes of almost all biomolecules are extremely rugged, having numerous local minima separated by barriers of different heights.
Put another way, free energy basins in this landscape can typically be subdivided into smaller local minima, giving rise to a hierarchy of minima.
Transitions across low barriers occur exponentially more often than those across higher barriers.
Groups of local minima separated by low barriers will mix rapidly and, therefore, appear as a single larger state to other minima separated from them by larger barriers.
Thus, these groups can satisfy a requirement for coarse-graining models called lumpability \cite{lumpability}.
A microstate MSM is considered lumpable with respect to some set of macrostates if and only if, for every pair of macrostates $M_1$ and $M_2$ and any pair of microstates $i$ and $j$ in $M_1$
$\sum_{k\in{M_2}}p_{ik}
=\sum_{k\in{M_2}}p_{jk}$ where $p_{ij}$ is the probability that the system will transition to state $j$ given that it is currently in state $i$.

We can exploit the concept of lumpability to construct coarse-grained models by progressively lumping together the most kinetically similar states---i.e. those with similar transition probabilities.
Physically, this is equivalent to merging states that mix rapidly because they are only separated by a low free energy barrier. 
One might be tempted to use an L1 or L2 norm between the transition probabilities out of each pair of states to determine which are most similar.
However, such an approach would ignore the fact that some states and transitions are better sampled than others and, therefore, would be susceptible to the same pitfalls as eigenspectrum-based methods.

I propose a Bayesian method for determining which states to lump together.
Specifically, I propose to employ a Bayes factor comparing how likely the data observed for a pair of states is to have come from either different ($P(\textrm{different}|C)$) or the same ($P(\textrm{same}|C)$) underlying distribution of transition probabilities
\begin{align}
&\frac{P(\textrm{different}|C)}{P(\textrm{same}|C)}
\label{eq:genBayesFactor}
\end{align}
where $C$ is the matrix of transition counts observed between all pairs of states.
Bayes factors compare the evidence (or marginal likelihood, $P(\textrm{Model}|\textrm{Data})$) for two different models.
In calculating these marginal likelihoods, one integrates over all possible parameterizations of a model, thereby accounting for uncertainty.
Therefore, one can construct a hierarchy of coarse-grained models in a manner that explicitly accounts for statistical uncertainty in a model by repeatedly calculating the BACE Bayes factor for every pair of states and then merging the two states with the smallest Bayes factor (i.e. the states that are most likely to have come from the same underlying distribution of transition probabilities).

A number of approximations are useful for making this approach computationally efficient.
For example, a brute force implementation of this algorithm where we recalculated every Bayes factor during each iteration of the algorithm would be quite inefficient, having a computational complexity of  $O(n^4)$.
We can achieve a complexity of $O(n^3)$---which is equivalent to eigenspectrum-based methods---by recognizing that merging two states has a negligible effect on Bayes factors not involving either of them and only recalculating Bayes factors including the new merged state.
We can also avoid a number of computations by only computing Bayes factors for connected states---i.e. pairs of states with at least one direct transition between them.
Disconnected state are likely to be separated by large free energy barriers, so they will necessarily have large Bayes factors and should not be merged.
Finally, it is valuable to derive an approximate expression for the BACE Bayes factor.
One could evaluate the Bayes factor by sampling from the posterior distribution for each state.
However, doing so would require a number of calculations for every comparison of a pair of states.
A single, closed-form expression---like the one derived in the next section---is significantly more efficient.

The BACE algorithm is then
\begin{enumerate}
\item Starting at the microstate level, calculate the BACE Bayes factor for every pair of connected states using the closed-form approximation from Eq. \ref{eq:bayesFact} derived in the next section.
\item Identify the pair of states with the smallest Bayes factor (i.e. the states that are most likely to have come from the same underlying distribution) and merge them by summing their transition counts.
\item Update the Bayes factors comparing the new merged state and every other state it is connected to, again using the approximate expression for the BACE Bayes factor from Eq. \ref{eq:bayesFact}.
\item Repeat steps 2 and 3 until only two states remain.
\end{enumerate}

One could also stop the algorithm when the BACE Bayes factor reaches a certain threshold.
For example, a $log_{10}(\textrm{Bayes factor})$ of 1 indicates that the model in the numerator is significantly more likely (over ten times more likely) than the one in the denominator.
Therefore, if the minimum BACE Bayes factor between any pair of states reaches 1, then one could infer the any further merging of states would greatly reduce the quantitative accuracy of the model and stop the algorithm.
However, if one's objective is to understand a system then continuing to merge states may be of great value.
The resulting models will only be qualitatively correct, at best.
However, their simplicity may allow more insight.
Hypotheses generated with these simple, qualitative models can then be tested with more complex, quantitative models and, ultimately, with experiments.

Future improvements to the algorithm could be made by including more complex moves.
For example, the current algorithm is greedy and, therefore, can never recover if two states are mistakenly merged.
One could correct such mistakes by allowing microstates to move to more appropriate macrostates or by iteratively breaking macrostates apart and rebuilding them, as in Ref. \cite{chodera}.
Such moves are not undertaken here as they would reduce the efficiency of the method.
Moreover, the present greedy algorithm performs quite well compared to other methods, as discussed in the Results section.

\section{BACE Bayes Factor}
To make the BACE algorithm computationally efficient, I derive a closed-form expression for the log of the BACE Bayes factor from Eq. \ref{eq:genBayesFactor}.
The final expression for the BACE Bayes factor is
\begin{align}
&\log{\frac{P(\textrm{different}|C)}{P(\textrm{same}|C)}} \approx \hat{C}_i\mathcal{D}(p_i\|q)+\hat{C}_j\mathcal{D}(p_j\|q)
\label{eq:bayesFact}
\end{align}
where $C$ is the transition count matrix, 
$\hat{C}_i$ is the number of transitions observed from state $i$, 
$\mathcal{D}(p_i\|q)=\sum_k{p_{ik}\log{\frac{p_{ik}}{q_k}}}$ is the relative entropy between probability distribution $p_i$ and $q$, 
$p_i$ is a vector of maximum likelihood transition probabilities from state $i$, and $q=\frac{\hat{C}_i p_i+\hat{C}_j p_j}{\hat{C}_i+\hat{C}_j}$ is the vector of expected transition probabilities from combining states $i$ and $j$.
Note that this expression includes a comparison between $p_{ij}$ and $q_j$ that helps prevent the merger of disconnected states.
For example, consider the simple model $A \leftrightarrow B \leftrightarrow C$ ($A$ and $C$ are disconnected).
If the BACE Bayes factor only compared transition probabilities to states other than the two being considered, then one could easily obtain lumpings like $\{A,C\}$ and $\{B\}$.
However, comparing the self-transition probabilities and exchange probabilities between the states being compared helps avoid these pathological situations (e.g. in this case $p_{AA}>0$ while $p_{CA}=0$, so these states are unlikely to appear kinetically close).

This expression is equivalent to the generalized Jesnen-Shannon divergence \cite{Jensen} and, therefore, has an appealing information theoretic interpretation.
Given a sample drawn from one of two probability distributions, the Jensen-Shannon divergence is the average information that sample provides about the identity of the distribution it was drawn from \cite{Jensen2}.
The result is zero if the two distributions are equivalent and reaches its maximal value if the distributions are non-overlapping and a single data point, therefore, uniquely specifies which distribution it was drawn from.
In this case, the larger the Bayes factor is, the more likely the data for each state are to have come from different underlying distributions.
By iteratively merging the most kinetically similar states, BACE retains the most divergent states, which can be interpreted as keeping the states with the most information content.

To derive Eq. \ref{eq:bayesFact}, we first recognize that every possible set of transition probabilities out of some initial state that satisfies $0 \leq \breve{p}_{ij} \leq 1$ and $\sum_j{\breve{p}_{ij}}=1$ has some probability of generating the observed transitions out of that state.
From Bayes rule, the posterior probability of some distribution ($\breve{p}_i$) being the true underlying distribution given a set of observed transitions is
\begin{align}
P(\breve{p}_i|C_i,\alpha_i) \propto P(C_i|\breve{p}_i)P(\breve{p}_i|\alpha_i)
\label{eq:bayesRule}
\end{align}
where $C_i$ is a vector of transition counts out of state $i$ and $\alpha_i$ will be discussed shortly.

Assuming the transition probabilities for each state are independent, we can use a multinomial distribution for the likelihood
\begin{align*}
P(C_i|\breve{p}_i)=\frac{\hat{C}_i!}{\prod_k{C_{ik}!}}\prod_k{\breve{p}_{ik}^{C_{ik}}}
\end{align*}

A Dirichlet prior ($D$) is chosen as it is conjugate to the multinomial likelihood.
That is, if the prior is a Dirichlet then the posterior is also a Dirichlet.
The prior is then
\begin{align*}
P(\breve{p}_i|\alpha_i)&=D(\alpha_i)=\frac{\Gamma(\sum_k{\alpha_{ik}})}{\prod_k{\Gamma(\alpha_{ik})}}\prod_k{\breve{p}_{ik}^{\alpha_{ik}-1}}
\end{align*}
where $\alpha_i$ is a vector of pseudocounts giving the expected number of transitions before any data is observed.
We choose $\alpha_{ik}=1/n$ where $n$ is the number of states because for a state to exist we must have observed at least one transition originating from that state and, prior to observing any data, the chance that that transition is to any particular state is equal \cite{Hinrichs:2007p1865,BOWMAN:2010p17}.

Combining the expressions for the likelihood and prior, the posterior distribution from Eq. \ref{eq:bayesRule} is
$
P(\breve{p}_i|C_i,\alpha_i)=D(C_i+\alpha_i)
$.

We can now calculate the log of the evidence for a particular model (M)
\begin{align}
\log{P(C_i|M)} &= \log\int_{\breve{p}_i}{P(C_i|\breve{p}_i)P(\breve{p}_i|\alpha_i)}	\notag \\
&\approx \log\frac{\Gamma(\sum_k{\alpha_{ik}})}
{\Gamma(\sum_k{[C_{ik}+\alpha_{ik}]})}
\prod_k{\frac{\Gamma(C_{ik}+\alpha_{ik})}{\Gamma(\alpha_{ik})}} \label{eq:gamma}
\end{align}
\begin{align}
&\approx \sum_k{C_{ik}\log{p_{ik}}} - n\log{n}+n	\notag\\
&\approx -\hat{C}_{i}\mathcal{H}(p_i) - n\log{n}+n \label{eq:evidence}
\end{align}
where $\mathcal{H}(p_i)=-\sum_k{p_{ik}\log{p_{ik}}}$ is the entropy of $p_i$ and we have made the substitutions $\hat{C}_i=\sum_k{C_{ik}}$, $p_{ik}=C_{ik}/\hat{C}_{i}$ (the maximum likelihood estimate of the transition probability), $\Gamma(C_{ik}+1/n) \approx \Gamma(C_{ik}+1)=C_{ik}!$, $\Gamma(1/n) \approx n$, and Stirling's approximation.
Note that the approximations made between Eqs. \ref{eq:gamma} and \ref{eq:evidence} break down for small sample sizes but this can be ignored safely.
One could calculate the evidence more accurately by directly evaluating Eq. \ref{eq:gamma}.
However, this could lead to numerical errors as the $\Gamma$ function tends to blow-up for the large inputs one is likely to encounter in real-world applications of this method.
Moreover, the closed-form expression for the Bayes factor based on Eq. \ref{eq:evidence} performs quite well in practice, as discussed in the Results section.

The BACE Bayes factor given in Eq. \ref{eq:bayesFact} is then the ratio of the evidence for the 
transition counts from states $i$ and $j$ coming from two different distributions versus a single distribution
($\log{\frac{P(\textrm{different}|C)}{P(\textrm{same}|C)}}=\log{\frac{P(C|\textrm{different})P(\textrm{different})}{P(C|\textrm{same})P(\textrm{same})}}
$)
where we assume the prior probabilities for the two models are equal and drop terms depending only on $n$ as they simply introduce a constant that has no effect on the relative ordering of Bayes factors comparing various states.

\section{Approximate Bayesian Model Comparison}
Bayesian model comparison is a powerful way to determine which of two models best explains a set of observations.
Such methods are of great value here as they can be used to compare the results of BACE to other coarse-graining methods.
Moreover, they can be used to decide which levels of the hierarchy of models from BACE are most deserving of further analysis.
However, current methods \cite{Bacallado:2009p1009} are too computationally demanding for this second task.

Using similar mathematical machinery to that employed in the derivation of BACE and paralleling the derivation in Ref. \cite{Bacallado:2009p1009}, we can also derive a closed-form expression for the log of the Bayes factor comparing two coarse-grainings---or lumpings---of an MSM, $L_1$ and $L_2$,
\begin{align}
\log{\frac{P(L_1|C)}{P(L_2|C)}} \approx &\sum_{M \in L_2}{\hat{B}_M[\mathcal{H}(p_M)+\mathcal{H}(\Theta_M)]}	\notag \\
&-\sum_{M \in L_1}{\hat{B}_M[\mathcal{H}(p_M)+\mathcal{H}(\Theta_M)]}
\label{eq:modelComparison}
\end{align}
where $B$ and $C$ are the transition count matrices at the macrostate and microstate levels, respectively, $M$ is a macrostate in lumping $L$, $\hat{B}_M$ is the number of transitions originating in $M$, $p_M$ is a vector of the maximum likelihood transition probabilities from $M$, $\Theta_M$ is a vector of the maximum likelihood probabilities of being in each microstate $m$ given that the system is in $M$, and $\mathcal{H}$ is the entropy.
Evaluating this expression is extremely efficient, making it feasible to compare the merits of each model in the hierarchy generated by BACE.

To derive the expression for model comparison from Eq. \ref{eq:modelComparison}, we need to calculate the evidence for a particular coarse-graining, $L$, 
\begin{align*}
\log{P(C|L)} = \log{\int_T{\int_\Theta{P(B|T,L)P(C|B,\Theta,L)P(T,\Theta)}}}
\end{align*}
where $T$ is the macrostate transition probability matrix.
Because the macrostate trajectory and selection of microstates are independent, this can be rewritten as
\begin{align}
\log{P(C|L)} = \log{\int_T{P(B|T,L)P(T)}}+\log{\int_\Theta{P(C|B,\Theta,L)P(\Theta)}}
\label{eq:modelEvidence}
\end{align}

Assuming the transition counts from each state come from independent multinomial distributions and using similar reasoning as employed in the derivation of BACE, the first term in Eq. \ref{eq:modelEvidence} is
\begin{align*}
\log{\int_T{P(B|T,L)P(T)}} \approx -\sum_{M \in L}{\hat{B}_M\mathcal{H}(p_M)}
\end{align*}
From Ref. \cite{Bacallado:2009p1009}, the second term in the expression for model comparison from Eq. \ref{eq:modelComparison} is
\begin{align*}
\log{\int_\Theta{P(C|B,\Theta,L)P(\Theta)}} \approx \log{\prod_{M \in L}{
\frac{\Gamma(|M|)\prod_{m \in M}{\Gamma(\hat{C}_m+1)}}
{\Gamma(\hat{B}_M+|M|)}
}}
\end{align*}
where $m$ is a microstate in macrostate $M$, $|M|$ is the number of microstates in $M$,  and we have assumed a pseudocount of 1 to reflect our prior belief that for a microstate to exist, we must have observed at least one transition originating in that state.
Using $\frac{\Gamma(Y)}{\Gamma(X+Y)} \approx \frac{1}{X!}$ and, again, the reasoning from BACE, this becomes
\begin{align*}
\log{\int_\Theta{P(C|B,\Theta,L)P(\Theta)}} \approx -\sum_{M \in L}{\hat{B}_M\mathcal{H}(\Theta_M)}
\end{align*}

\section{Results}
BACE is much better at dealing with statistical uncertainty in model parameters than current eigenspectrum-based methods.
For example, it is able to correctly identify the three macrostates in the simple model shown in Fig. \ref{fig:simpleModel} even in the presence of the poorly sampled transitions that confound eigenspectrum-based methods.
BACE also naturally lumps states with few samples into larger ones, whereas eigenspectrum-based methods tend to make such states into singleton macrostates.
With BACE, a significantly better sampled state will dominate the Bayes factor when compared to a poorly sampled state, leading to a high likelihood that the poorly sampled state will be absorbed into its better sampled neighbor.
Methods like super-level-set hierarchical clustering (SHC) \cite{Huang:2010p22} and the most probable path algorithm \cite{stock} also deal with poorly sampled states by merging them into larger states.
However, in these methods, a state is considered poorly sampled if its population is below some user-defined cutoff.
BACE offers the advantage of naturally identifying poorly sampled states without any reliance on user-defined input.

Beyond this qualitative improvement, a quantitative measure of model validity shows that coarse-grainings from BACE capture both the thermodynamics and kinetics of molecular systems better than existing eigenspectrum-based methods (Table \ref{tab:compare}).
To draw this conclusion, I first built models for each model system using BACE, PCCA, and PCCA+ with the same number of macrostates.
I then employed a Bayesian method for model comparison to determine which model is most consistent with the original data.
This method calculates a Bayes factor comparing the evidence for two different coarse-grainings while taking into account many of the constraints on valid MSMs, like reversibility \cite{Bacallado:2009p1009}.
It should not be confused with the BACE Bayes factor, which compares two states.
If the values from the Bayesian model comparison algorithm are large ($>1$), then the model in the numerator is significantly more likely to have generated the observed data than the model in the denominator, whereas the model in the denominator is better if these values are small ($<-1$).
Intermediate values (between $1$ and $-1$) suggest that neither model is strongly preferred over the other. 
Both this model comparison method and the approximate version outlined here quantitatively compare the consistency of two coarse-grainings with the original microstate trajectories.
This comparison integrates over all possible macrostate transition probability matrices and all possible microstate equilibrium probabilities within each macrostate for each coarse-graining and, therefore, captures both the thermodynamics and kinetics of each model.
Table \ref{tab:compare} shows that  BACE is typically many orders of magnitude better than eigenspectrum-based methods by this metric.
Such quantitative comparisons are crucial because the complexity of most real-world MSMs renders a qualitative assessment of a coarse-graining's validity impossible.

\begin{table}
\caption{\label{tab:compare}
Comparison of BACE with eigenspectrum-based methods for a series of model systems from the simple model shown in Fig. \ref{fig:simpleModel} to a full protein, the villin headpiece.
The same number of states is used for each method.
The numbers reported are the $log_{10}$ of the Bayes factor comparing how likely the coarse-graining from BACE is to have generated a given data set to how likely the model from PCCA (or PCCA+) is to have generated the data.
Numbers greater than one suggest the model from BACE is significantly better at explaining the observed data than the models from other methods whereas numbers less than -1 would suggest the other methods are significantly better than BACE.
These values were calculated with the model comparison method from Ref \cite{Bacallado:2009p1009} with 100 bootstrapped samples.
Mean and $68\%$ confidence interval are reported.
The large numbers are comparable to those found in Ref. \cite{Bacallado:2009p1009} and arise from the products of a large number of small probabilities in the likelihood function.
The zero entry for comparing the performance of BACE and PCCA+ on the simple model arises from the fact that they give equivalent results in this case.
}
\begin{ruledtabular}
\begin{tabular}{lcc}
Model & $\log_{10}\frac{P(BACE|C)}{P(PCCA|C)}$ & $\log_{10}\frac{P(BACE|C)}{P(PCCA+|C)}$   \\
\hline
Simple$^a$ & $1324\ (1079,1548)$ & $0$ \\
Alanine dipeptide$^b$ & $3239\ (3152,3312)$ & $2707\ (2573,2862)$ \\
Villin$^c$ & $11450\ (10913,12038)$ & $16997\ (16076,17856)$ \\
\end{tabular}
\end{ruledtabular}
\raggedright$^a$ Model with 9 microstates and 3 macrostates from Fig. \ref{fig:simpleModel}.

$^b$ Model with 181 microstates and 6 macrostates from Ref. \cite{Beauchamp:2012p2556}.

$^c$ Model with 10,000 microstates and 500 macrostates from Ref. \cite{BOWMAN:2009p10}.
\end{table}

\begin{figure}
\includegraphics[scale=0.8]{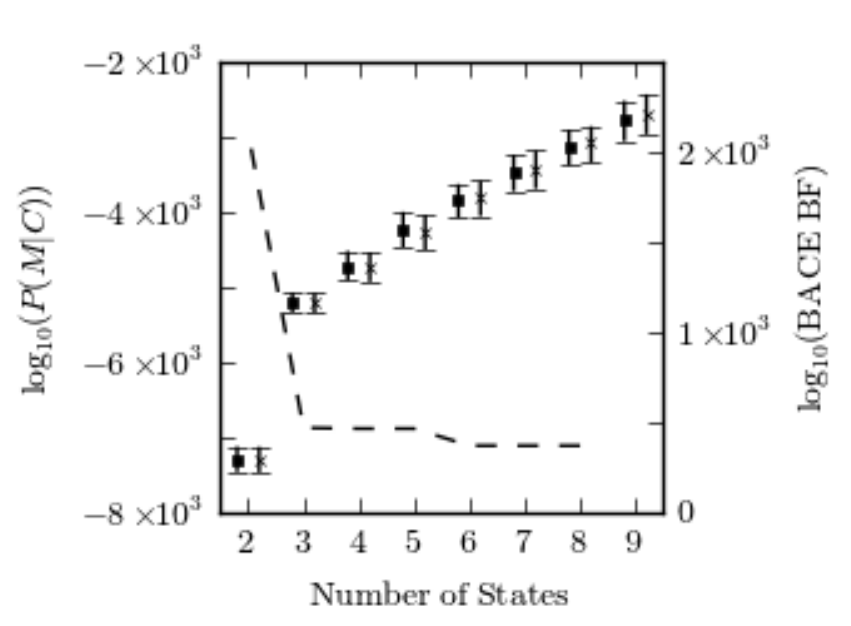}
\caption{\label{fig:bayesFactor} 
The evolution of the Bayes factors as the states from the simple model in Fig. \ref{fig:simpleModel} are progressively merged together indicates the most meaningful levels of the hierarchy of models.
The BACE Bayes factor (BACE BF) is plotted as a dashed line (values on right axis).
The means and $68\%$ confidence intervals of the evidence from the approximate model comparison expression (asterisks) and the more exact method enforcing reversibility (squares) are also plotted (values on left axis).
Drastic changes occur in all three curves when kinetically distinct states are merged.
Models immediately preceding these costly mergers are likely good candidates for further analysis as they contain a maximal amount of information with a minimal number of states.
A second important point is that the approximate Bayes factor for model comparison derived here tracks the more exact expression well in this case.
Therefore, the more computationally efficient approximation can be used in place of the more exact but costly expression.
}
\end{figure}

\begin{figure}
\includegraphics[scale=0.8]{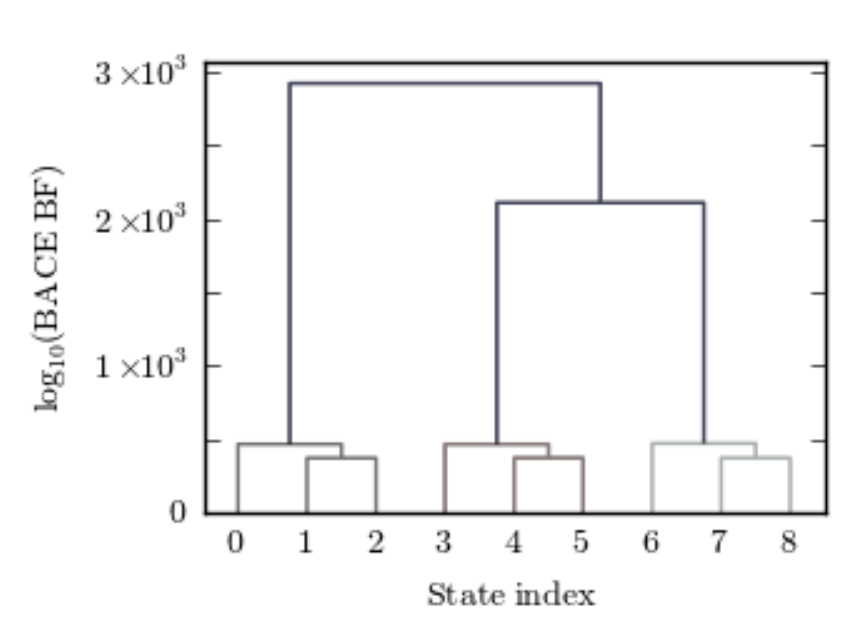}
\caption{\label{fig:dendro} 
A dendrogram representation of the BACE Bayes factors from the simple model in Fig. \ref{fig:simpleModel} captures the hierarchical nature of the underlying landscape.
The states are numbered from 0 to 8 on the x-axis.
The brackets connect states that are being merged and the y-values of the crossbars of these brackets are the BACE Bayes factors between the states being merged.
This representation highlights that the three kinetically similar microstates within each macrostate are merged together first (small Bayes factors).
Subsequent merger of the more kinetically dissimilar macrostates has a much greater cost (larger Bayes factors).
}
\end{figure}

Another advantage of BACE is that it generates an entire hierarchy of models.
Having this hierarchy makes it possible to look for general properties that are robust to the degree of coarse-graining and, therefore, may be important properties of the system being investigated.
In addition, having this hierarchy allows the user to determine how many macrostates are appropriate to use.
In theory, one could employ the Bayesian model comparison method accounting for reversibility from Ref. \cite{Bacallado:2009p1009} to decide which levels of the hierarchy are most deserving of further analysis but, in practice, this would be impractical due to the time requirements of that method.
However, both the BACE Bayes factor and the approximate model comparison method presented here correlate well with the reversible method (Fig. \ref{fig:bayesFactor}) and, therefore, can be used to guide which levels of the hierarchy are pursued further.
Each Bayes factor changes more rapidly when more distinct states are lumped together, so models immediately preceding these dramatic jumps are ideal for further analysis.
The BACE Bayes factor can even be used to visualize the hierarchical nature of a system's free energy landscape and choose appropriate levels for further analysis (Fig. \ref{fig:dendro}).
One could also combine the methods by using the approximate expression to guide the application of the reversible method.

\section{Conclusions}
I have presented a Bayesian agglomerative clustering engine (BACE) for coarse-graining MSMs that significantly outperforms existing methods in capturing the thermodynamics and kinetics of molecular systems.
The bottom-up nature of the algorithm likely makes it especially well suited for constructing mesoscale models.
The method is also directly applicable to other Markov chains and could easily be extended to other probabilistic models.

The development of the method was guided by physical intuition regarding the hierarchical nature of the free energy landscapes that ultimately govern the structure and dynamics of molecular systems.
The final result is equivalent to the generalized Jensen-Shannon divergence, giving the method an appealing information theoretic interpretation in terms of the information content of a measurement.
Therefore, BACE could greatly facilitate a deeper understanding of molecular systems.
In particular, it can provide an entire hierarchy of models that captures the hierarchical nature of a molecule's free energy landscape.
The Bayes factors derived here can be used to guide which levels of the hierarchy are used for analysis and a fast, approximate expression for model comparison derived here may prove valuable in situations where more exact expressions are too expensive.

\section{Simulation Details}
The alanine dipeptide data used for Table \ref{tab:compare} was taken from Ref. \cite{Beauchamp:2012p2556}.
One hundred simulations were performed with Gromacs 4.5 \cite{gromacs} using the Amber96 force field \cite{amber96} and the OBC GBSA implicit solvent \cite{obc}.
Each trajectory is $500 ps$ long, with conformations stored every $1 ps$.

The villin data used for Table \ref{tab:compare} was taken from Ref. \cite{villin} and the macrostate definitions were taken from Ref. \cite{BOWMAN:2009p10}.
Five hundred simulations were performed with Gromacs deployed on the Folding@home distributed computing environment \cite{villin,fah}.
The Amber03 force field \cite{amber03} and Tip3p explicit solvent were used.
Each trajectory is up to $2 {\mu}s$ long, with conformations stored every $50 ps$.

\begin{acknowledgments}
I am grateful to the Miller Institute for funding and to TJ Lane and DA Sivak for helpful comments.
\end{acknowledgments}


\end{document}